\documentclass[pra,superscriptaddress,twocolumn]{revtex4-2}

\usepackage{enumerate}
\usepackage{amsfonts,amssymb,amsmath}
\usepackage[]{graphics,graphicx,epsfig}
\usepackage{amsthm}
\usepackage{graphicx}
\usepackage{dcolumn}
\usepackage{natbib}
\usepackage{color}
\usepackage{multirow}
\usepackage{ulem}
\usepackage{bm}
\usepackage{url}
\usepackage{braket}
\usepackage[caption=false]{subfig}

\newcommand{\measA}{\mathfrak{A}}
\newcommand{\measB}{\mathfrak{B}}

\begin{document}


\title{Bi-Contextuality: A Novel Non-Classical Phenomenon in Bipartite Quantum Systems}

\author{Gabriel Ruffolo}
\affiliation{Instituto de F\'{i}sica Gleb Wataghin, Universidade Estadual de Campinas (Unicamp),
Rua S\'{e}rgio Buarque de Holanda 777, Campinas, S\~{a}o Paulo 13083-859, Brazil}

\author{Nigel Benjamin Lee Junsheng}
\affiliation{Centre for Quantum Technologies,
National University of Singapore, 3 Science Drive 2, 117543 Singapore,
Singapore}

\author{Kim Mu Young}
\affiliation{Centre for Quantum Technologies,
National University of Singapore, 3 Science Drive 2, 117543 Singapore,
Singapore}

\author{Dzmitry Matsukevich}
\affiliation{Centre for Quantum Technologies,
National University of Singapore, 3 Science Drive 2, 117543 Singapore,
Singapore}
\affiliation{Department of Physics,
National University of Singapore, 3 Science Drive 2, 117543 Singapore,
Singapore}

\author{Rafael Rabelo}
\affiliation{Instituto de F\'{i}sica Gleb Wataghin, Universidade Estadual de Campinas (Unicamp),
Rua S\'{e}rgio Buarque de Holanda 777, Campinas, S\~{a}o Paulo 13083-859, Brazil}

\author{Dagomir Kaszlikowski}
\email{phykd@nus.edu.sg}
\affiliation{Centre for Quantum Technologies,
National University of Singapore, 3 Science Drive 2, 117543 Singapore,
Singapore}
\affiliation{Department of Physics,
National University of Singapore, 3 Science Drive 2, 117543 Singapore,
Singapore}

\author{Pawe{\l} Kurzy{\'n}ski}
\email{pawel.kurzynski@amu.edu.pl}
\affiliation{ Institute of Spintronics and Quantum Information, Faculty of Physics, Adam Mickiewicz University, Uniwersytetu Pozna{\'n}skiego 2, 61-614 Pozna\'n, Poland}


\date{\today}


\begin{abstract}
We present and experimentally demonstrate a novel non-classical phenomenon, bi-contextuality, observed in quantum systems prepared by two independent sources. This discovery plays a key role in the developing framework of network nonlocality, offering a new method for confirming the quantum nature of measurements within a single network node. Bi-contextuality acts as a reversed Bell scenario: while Bell scenarios involve splitting a system for independent measurements, our approach combines systems from separate independent sources for joint measurements. The outcomes defy classical models that assume independence and non-contextuality. The simplest Bell scenario can be seen as a subset of the Peres-Mermin (PM) square, and our phenomenon represents another important subset of this framework. Moreover, bi-contextuality has notable consequences related to the Pusey-Barrett-Rudolph (PBR) theorem, suggesting that classical psi-ontic models must account for contextuality or the dependence of preparations, challenging established assumptions. 


\end{abstract}

\maketitle


Quantum mechanics challenges classical intuition, with phenomena such as superposition, entanglement, and non-locality revealing the inherently non-classical nature of quantum systems. 
While these phenomena have been extensively studied in single-source systems \cite{brunner2014,budroni2022}, the behavior of quantum systems within larger networks has recently gained attention \cite{tavakoli2022, branciard2012, gisin2017, abiuso2022}. In this context, the concept of nonlocality has been extended to quantum networks, where multiple nodes are involved in the measurement and processing of quantum information. These networks provide a platform for exploring the boundaries between classical and quantum physics, and have practical implications for quantum communication and computing.

A key concept in quantum theory is contextuality, the idea that the outcomes of quantum measurements cannot always be described by classical models that assume measurement outcomes are independent of other factors, such as the measurement context \cite{budroni2022,liang2011specker, kochen1990problem}. This concept has been central in discussions of nonlocality, particularly in the Bell scenario, which involves measurements of two parts of a single system that are spatially separated. Bell’s Theorem and its variants show that such systems cannot have a local and realistic description \cite{brunner2014}.

In this paper, we introduce and experimentally demonstrate a novel non-classical phenomenon, called bi-contextuality, in bipartite quantum systems prepared by two independent sources. Unlike Bell-type experiments, where parts of a single system are separated and measured independently, we consider systems originating from separate, individual sources that are combined for joint measurements. This approach challenges classical models that assume measurement outcomes stem from individual properties of multiple independent sources.

The phenomenon of bi-contextuality plays an interesting role in the emerging framework of network nonlocality. In particular, it offers a new method for certifying the quantum nature of measurements in a single node of a network, where the independence of the sources and the non-contextuality of the classical model are key assumptions. This extension of network nonlocality opens new avenues for verifying the quantum behavior of complex systems.

Moreover, we show that bi-contextuality is linked to the Peres-Mermin (PM) square \cite{peres1990, mermin1990}, providing a new subset of this well-established scenario, alongside the Bell scenario. We also demonstrate that bi-contextuality has important implications related to the Pusey-Barrett-Rudolph (PBR) theorem \cite{pusey2012}, which asserts that any classical model describing quantum states must adhere to strict conditions on the overlap between probability distributions over hypothetical ontic states for different quantum states \cite{leifer2014}. Our work suggests that these models must account for contextuality or the dependence on preparation procedures.

\section{Results}

\subsection{Preliminaries}

In a quantum network scenario, there are independent sources, denoted by $\lambda_i$ $(i=1,2,\ldots)$, sending systems to laboratories, denoted by $L=\mathcal{A},\mathcal{B},\ldots$. Sources that prepare systems to more than one laboratory can, in principle, distribute non-classical correlations (entanglement) \cite{tavakoli2022}. At each laboratory $X$, an experimenter measures all received systems and obtains an outcome $X$. Such measurement can be either fixed or randomly chosen from a set of available measurements. The goal of such a scenario is to determine whether the obtained measurement outcomes admit a classical local description. If they don't, we assert the quantumness of such distributed resources, which are referred to as {\it network nonlocality} \cite{tavakoli2022, renou2022, cavalcanti2011quantum, gu2023experimental}. These resources can be later used for various quantum communication and computation protocols \cite{pironio2010random, acin2007device, acin2016certified}

\begin{figure}[t]
\includegraphics[width=0.35\textwidth]{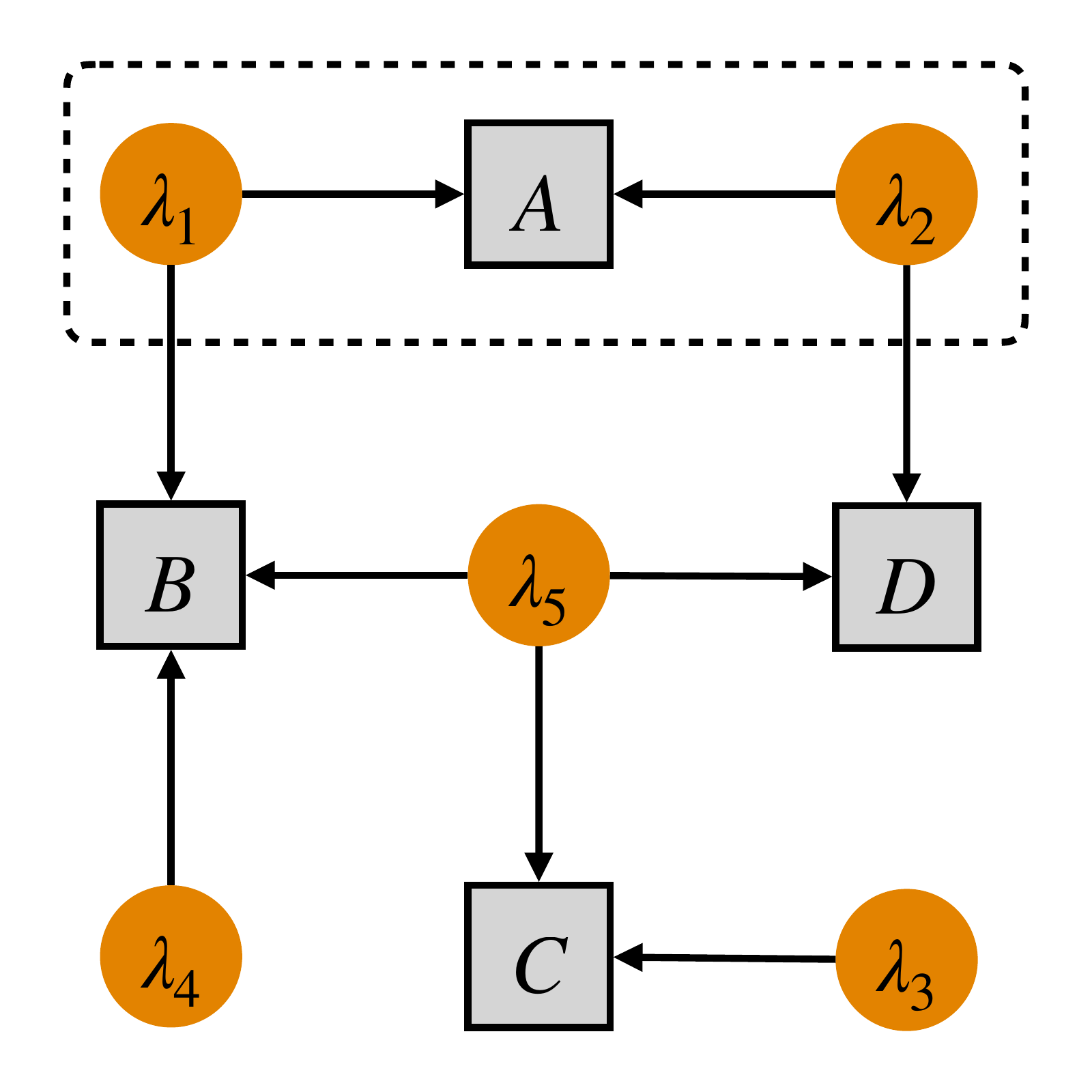}
\vspace{-5mm}
\caption{An example of a network scenario with five independent sources $\lambda_i$ $(i=1,\ldots,5)$ and four laboratories $L=\mathcal{A},\mathcal{B},\mathcal{C},\mathcal{D}$. The subset of this scenario, inside the dashed border, corresponds to a network contextuality scenario (in this case bi-contextuality, due to two sources).} 
\label{f0}
\end{figure}

It is instructive to consider an example of the above network scenario, see Fig. \ref{f0}. In this case, we have five sources, $\lambda_i$ $(i=1,\ldots,5)$ and four laboratories $L=\mathcal{A},\mathcal{B},\mathcal{C},\mathcal{D}$. If we assume that sources are classical, they generate systems that are described by classical probability distributions $\mu_i(\lambda_i)$. In addition, if we assume that measurement outcomes are generated in a classical way, then the corresponding probabilities of outcomes are conditioned on the sources. For example, the measurement outcome A, obtained in laboratory $\mathcal{A}$, depends only on $\lambda_1$ and $\lambda_2$. Therefore, the probability distribution of the corresponding outcome is
\begin{equation}
p(A)=\int d\lambda_1 \int d\lambda_2 \Big(\mu_1(\lambda_1) \mu_2(\lambda_2) p(A|\lambda_1,\lambda_2)\Big),  \label{networkcont}  
\end{equation}
where we explicitly assumed independence of sources, i.e., the probability distribution of both sources is described by a product $\mu(\lambda_1,\lambda_2)=\mu_1(\lambda_1)\mu_2(\lambda_2)$. Probability distributions of remaining outcomes, $p(B)$, $p(C)$, and $p(D)$ can be obtained in an analogous way. Finally, a joint distribution of outcomes is given by
\begin{eqnarray}
& &p(A,B,C,D)=\int d\lambda_1 \int d\lambda_2 \int d\lambda_3 \int d\lambda_4 \int d\lambda_5 \times \nonumber \\
& &\Big( \mu_1(\lambda_1) \mu_2(\lambda_2)  \mu_3(\lambda_3) \mu_4(\lambda_4)\mu_5(\lambda_5) p(A|\lambda_1,\lambda_2)  \times \nonumber \\
& & p(B|\lambda_1,\lambda_4,\lambda_5) p(C|\lambda_3,\lambda_5) p(D|\lambda_2,\lambda_5) \Big).  
\label{clas}
\end{eqnarray}
The goal is to determine whether the observable distributions $p(A,B,C,D)$ admit a classical distribution (\ref{clas}). If they do not, we witness network nonlocality. 

The idea of this work is to focus on a single node of such a network where measurements on systems coming from many independent sources are performed. It is presented in Fig. \ref{f0} inside the dashed border. Since scenarios with measurements in a single part are typically known as contextuality scenarios \cite{budroni2022}, we call their network versions, which include many independent sources, network contextuality scenarios. We are going to consider the most fundamental network contextuality scenario with two independent sources, which we call {\it bi-contextuality}.


\subsection{Bi-Contextuality}

\begin{figure}[t]
\includegraphics[width=0.5\textwidth]{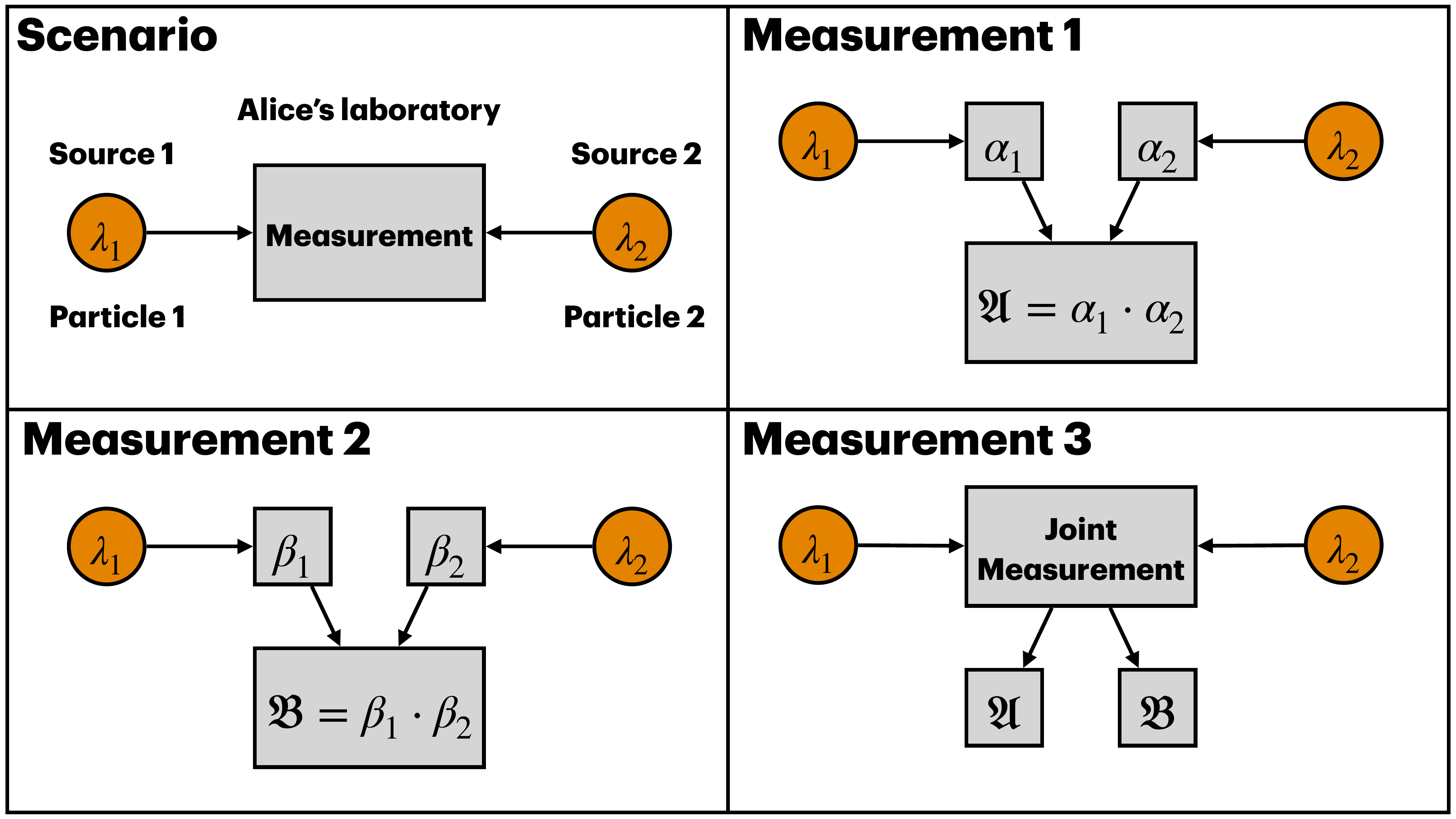}
\caption{Schematic representation of bi-contextuality scenario. Alice receives systems from two independent sources, $\lambda_1$ and $\lambda_2$. She chooses one of three possible measurement settings. In the first one, she separately measures $\alpha_1$ and $\alpha_2$ and then she evaluates their product $\measA$. Similarly, in the second one, she separately measures $\beta_1$ and $\beta_2$, and evaluates their product $\measB$. In the third one, she jointly measures both systems and evaluates $\measA$ and $\measB$, but not $\alpha_i$ and $\beta_{i}$.} 
\label{f1}
\end{figure}

We consider a single laboratory where an experimenter, Alice, performs measurements on systems coming from two independent sources $\lambda_1$ and $\lambda_2$. The goal is to determine if such measurements admit a classical description (\ref{networkcont}). In particular, our scenario consists of four individual properties, $\alpha_i,\beta_i = \pm 1$, that depend solely on $\lambda_i$ ($i=1,2$), and two bipartite properties $\measA, \measB= \pm 1$, where $\measA=\alpha_1\cdot\alpha_2$ and $\measB=\beta_1\cdot\beta_2$, that depend on both, $\lambda_1$ and $\lambda_2$. More specifically, we consider three different measurement settings. In the first setting, Alice measures $\alpha_1$, $\alpha_2$, and their product $\measA$. In the second setting, she measures $\beta_1$, $\beta_2$, and their product $\measB$. In the last setting, she performs a joint measurement on both systems to learn the values of products $\measA$ and $\measB$, but she does not learn individual values of $\alpha_i$ and $\beta_i$. This scenario is schematically represented in Fig. \ref{f1}.

The above scenario allows Alice to evaluate the following probability distributions: 
\begin{equation}
p(\alpha_1,\alpha_2),~~p(\beta_1,\beta_2),~~p(\measA,\measB).   
\end{equation}
Note that $p(\alpha_1,\alpha_2)$ and $p(\beta_1,\beta_2)$ allow evaluation of $p(\measA)$ and $p(\measB)$, respectively. This is because $p(\measA=+1)=p(\alpha_1=+1,\alpha_2=+1)+p(\alpha_1=-1,\alpha_2=-1)$, and so on. In addition, Alice can evaluate averages: 
\begin{eqnarray}
& & \langle \alpha_1 \rangle,~\langle \alpha_2 \rangle,~\langle \measA\rangle=\langle \alpha_1 \alpha_2\rangle=\langle \alpha_1 \rangle\langle\alpha_2\rangle, \\ & &\langle \beta_1 \rangle,~\langle \beta_2 \rangle,~\langle \measB\rangle=\langle \beta_1 \beta_2\rangle = \langle \beta_1\rangle \langle \beta_2\rangle, \\ & & \langle \measA\rangle,~\langle \measB\rangle,~\langle \measA \measB \rangle.
\end{eqnarray}
We assume that the non-disturbance assumption holds \cite{budroni2022}, i.e.,
\begin{eqnarray}
    & & \sum_{\alpha_1,\alpha_2|\alpha_1\cdot\alpha_2=a}p(\alpha_1,\alpha_2) = \sum_{\measB=\pm 1}p(\measA=a,\measB), \\
    & & \sum_{\beta_1,\beta_2|\beta_1\cdot\beta_2=b}p(\beta_1,\beta_2) = \sum_{\measA=\pm 1}p(\measA,\measB=b).
\end{eqnarray}
Finally, observe that due to the independence of sources, the following must hold:
\begin{eqnarray}
 & &p(\alpha_1,\alpha_2)=p(\alpha_1)p(\alpha_2), \\
 & &p(\beta_1,\beta_2)=p(\beta_1)p(\beta_2),
\end{eqnarray}
and, as a consequence,
\begin{equation}
\langle \measA \rangle = \langle \alpha_1\rangle \langle \alpha_2 \rangle, ~~~~ \langle \measB \rangle = \langle \beta_1\rangle \langle \beta_2 \rangle.   
\end{equation}
However, in general,
\begin{equation}
\langle \measA \measB \rangle \neq \langle \alpha_1\rangle \langle \alpha_2 \rangle\langle \beta_1\rangle \langle \beta_2 \rangle,   
\end{equation}
because there might exist unmeasurable correlations
\begin{equation}
\langle \alpha_i \beta_i \rangle \neq \langle \alpha_i \rangle \langle \beta_i \rangle.
\end{equation}


\subsection{Non-contextual model}

The standard non-contextual model assumes that, instead of two independent variables $\lambda_1$ and $\lambda_2$, the sources are described by a joint variable $\lambda$. As a consequence,
\begin{eqnarray}
    p(\alpha_1,\alpha_2)= \int d\lambda \mu(\lambda) p(\alpha_1|\lambda)p(\alpha_2|\lambda),
\end{eqnarray}
and similar for $p(\beta_1,\beta_2)$ and $p(\measA,\measB)$. Due to Fine-Abramsky-Brandenburger Theorem \cite{fine1982, abramsky2011sheaf}, the non-contextual model is equivalent to the existence of a joint probability distribution (JPD) $p(\alpha_1,\alpha_2,\beta_1,\beta_2)$. Even though $\measA$ ($\measB$) can be determined from $\alpha_1$ and $\alpha_2$ ($\beta_1$ and $\beta_2$), in this case it is useful to introduce probability distributions 
\begin{eqnarray}
 & & p(\alpha_1,\alpha_2,\measA)=p(\alpha_1,\alpha_2)\delta_{\measA,\alpha_1\cdot\alpha_2}, \\
 & & p(\beta_1,\beta_2,\measB)=p(\beta_1,\beta_2)\delta_{\measB,\beta_1\cdot\beta_2},
\end{eqnarray}
where $\delta_{x,j}$ is the Kronecker delta, and to look for a JPD $p(\alpha_1,\alpha_2,\beta_1,\beta_2,\measA,\measB)$. Interestingly, the above JPD can always be constructed from the measurable distributions $p(\alpha_1,\alpha_2,\measA)$, $p(\beta_1,\beta_2,\measB)$, and $p(\measA,\measB)$, using the methods discussed in \cite{ramanathan2012}. In particular, \begin{equation}
p(\alpha_1,\alpha_2,\beta_1,\beta_2,\measA,\measB) = \frac{p(\alpha_1,\alpha_2,\measA) p(\beta_1,\beta_2,\measB)p(\measA,\measB)}{p(\measA)p(\measB)}.
\end{equation}
Note that this JPD recovers measurable marginals, i.e.,
\begin{eqnarray}
& & p(\alpha_1,\alpha_2,\measA)=\sum_{\beta_1,\beta_2,\measB}p(\alpha_1,\alpha_2,\beta_1,\beta_2,\measA,\measB), \\
& & p(\beta_1,\beta_2,\measB)=\sum_{\alpha_1,\alpha_2,\measA}p(\alpha_1,\alpha_2,\beta_1,\beta_2,\measA,\measB), \\
& & p(\measA,\measB)=\sum_{\alpha_1,\alpha_2,\beta_1,\beta_2}p(\alpha_1,\alpha_2,\beta_1,\beta_2,\measA,\measB).
\end{eqnarray}
The above implies that such a scenario is always non-contextual. 


\subsection{Condition for bi-contextuality}

Now we assume independence of $\lambda_1$ and $\lambda_2$, which implies that classical descriptions of measurable probability distributions are of the form
\begin{eqnarray}
    p(\alpha_1,\alpha_2) = p(\alpha_1)p(\alpha_2) &=& \int d\lambda_1  \mu_1(\lambda_1) p(\alpha_1|\lambda_1) \times \nonumber \\
    & &\int d\lambda_2\mu_2(\lambda_2)p(\alpha_2|\lambda_2),
\end{eqnarray}
\begin{eqnarray}
    p(\beta_1,\beta_2) = p(\beta_1)p(\beta_2) &=&  \int d\lambda_1  \mu_1(\lambda_1) p(\beta_1|\lambda_1) \times \nonumber \\ 
    & &\int d\lambda_2 \mu_2(\lambda_2)p(\beta_2|\lambda_2), 
\end{eqnarray}
\begin{eqnarray}
    & &p(\measA,\measB)=  \\ & &\int d\lambda_1 \int d\lambda_2 \Bigg( \mu_1(\lambda_1)\mu_2(\lambda_2) p(\measA|\lambda_1,\lambda_2)p(\measB|\lambda_1,\lambda_2)\Bigg). \nonumber
\end{eqnarray}
Independence and classicality also imply probability distributions $p(\alpha_i,\beta_i)$, not measurable in this scenario, that are of the form
\begin{eqnarray}
p(\alpha_i,\beta_i)= \int d\lambda_i \mu_i(\lambda_i) p(\alpha_i|\lambda_i)p(\beta_i|\lambda_i). \label{bicontcondit}
\end{eqnarray}
This implies 
\begin{equation}
\langle \measA \measB\rangle = \langle \alpha_1 \alpha_2 \beta_1 \beta_2 \rangle = \langle \alpha_1  \beta_1 \rangle \langle \alpha_2 \beta_2 \rangle. \label{cond2}
\end{equation}

Now, consider a particular realization of our scenario in which $\langle \measA \measB \rangle = 0$. This and Eq.~(\ref{cond2}) implies that either $\langle \alpha_1  \beta_1 \rangle=0$, or $\langle \alpha_2 \beta_2 \rangle = 0$. Next, observe that the probability distribution of two binary $\pm 1$ random variables can always be expressed as
\begin{equation}\label{bidistro}
    p(\alpha_i,\beta_i)=\frac{1}{4}\Big(1+\alpha_i\langle \alpha_i\rangle + \beta_i\langle \beta_i\rangle + \alpha_i\beta_i\langle \alpha_i\beta_i\rangle  \Big),
\end{equation}
(see Methods), which, in our case, for at least one value of $i$ reduces to 
\begin{equation}\label{cond3}
    p(\alpha_i,\beta_i)=\frac{1}{4}\Big(1+\alpha_i\langle \alpha_i\rangle + \beta_i\langle \beta_i\rangle \Big).
\end{equation}
However, the above distribution is negative for $|\langle \alpha_i \rangle| +|\langle \beta_i \rangle| > 1$, in which case this scenario does not admit a non-bi-contextual model (\ref{bicontcondit}).


\subsection{Quantum bi-contextuality}

Now we consider two qubits coming from independent sources, for which the four individual properties correspond to Pauli observables:
\begin{equation}\label{m1}
\alpha_1=X\otimes \openone,~~ \alpha_2=\openone \otimes Y,~~ \beta_1=Y\otimes \openone,~~ \beta_2=\openone \otimes X,
\end{equation}
and the bipartite properties correspond to products 
\begin{equation}\label{m2}
\measA=X\otimes Y,~~\measB= Y\otimes X.
\end{equation}
Let the qubits be prepared in a state $|\psi\rangle \otimes |\psi\rangle$, where 
\begin{equation}
\langle\psi|X|\psi\rangle=\langle\psi|Y|\psi\rangle=\frac{1}{\sqrt{2}},~~\langle\psi|Z|\psi\rangle=0.   
\label{eqn:optimal_state}
\end{equation}
The above, and the properties of Pauli operators, imply 
\begin{eqnarray}
& &\langle \alpha_1 \rangle_Q = \langle \alpha_2 \rangle_Q = \frac{1}{\sqrt{2}}, ~\langle \measA \rangle_Q=\langle \alpha_1 \rangle \langle \alpha_2 \rangle =\frac{1}{2}, \label{A} \\
& & \langle \beta_1 \rangle_Q = \langle \beta_2 \rangle_Q = \frac{1}{\sqrt{2}}, ~\langle \measB \rangle_Q= \langle \beta_1 \rangle_Q \langle \beta_2 \rangle_Q =\frac{1}{2}, \label{B} \\
& & \langle \measA \measB \rangle_Q =  \langle Z \otimes Z \rangle_Q = \langle Z \rangle_Q \langle Z \rangle_Q = 0. \label{AB}
\end{eqnarray}
where the symbol $\langle \bullet \rangle_Q$ stands for a mean quantity obtained through quantum theory. The discussion in the previous section implies that either for $i=1$ or $i=2$ the presumed probability distribution $ p(\alpha_i,\beta_i)$ is negative,
\begin{equation}
     p_Q(\alpha_i=-1,\beta_i=-1)=\frac{1}{4}(1-\sqrt{2})<0.
\end{equation}
Therefore, the measurements on these two independently prepared qubits do not admit a non-bi-contextual description, hence they are bi-contextual, despite being non-contextual.


\subsection{Experimentally testable inequality}

Now we present an experimentally testable inequality, whose violation implies bi-contextuality. Its detailed derivation is provided in Methods. The system exhibits bi-contextuality if all of the following nonlinear inequalities are violated
\begin{eqnarray}
   & & L_2 \leq \frac{\langle \measA \measB \rangle}{L_1} \leq R_2, \label{eqn:exp_inequality_1st}\\
   & & L_2 \leq \frac{\langle \measA \measB \rangle}{R_1} \leq R_2, \\
   & & L_1 \leq \frac{\langle \measA \measB \rangle}{R_2} \leq R_1, \\
   & & L_1 \leq \frac{\langle \measA \measB \rangle}{L_2} \leq R_1,
    \label{eqn:exp_inequality}
\end{eqnarray}
where
\begin{eqnarray}
     L_i &=& |\langle\alpha_i \rangle + \langle\beta_i \rangle|-1,  \\
     R_i &=& 1- |\langle\alpha_i \rangle - \langle\beta_i \rangle|.
 \end{eqnarray}
 The above condition is both sufficient and necessary. Therefore, if at least one of the above inequalities is not violated, we know that the system is non-bi-contextual and the corresponding model exists (see Methods). All four inequalities can be succinctly reduced to a single inequality, 
 \begin{equation}
 \big(\langle \measA \measB\rangle-min\big)\big(\langle \measA \measB\rangle-max\big)\leq 0,
 \label{nonlinear}
 \end{equation}
where $min, max$ are a minimum and maximum of four numbers: $L_1 L_2,L_1R_2,R_1L_2$, and $R_1R_2$.


\begin{figure}[!t]
\subfloat[]{\includegraphics[width=0.4\textwidth]{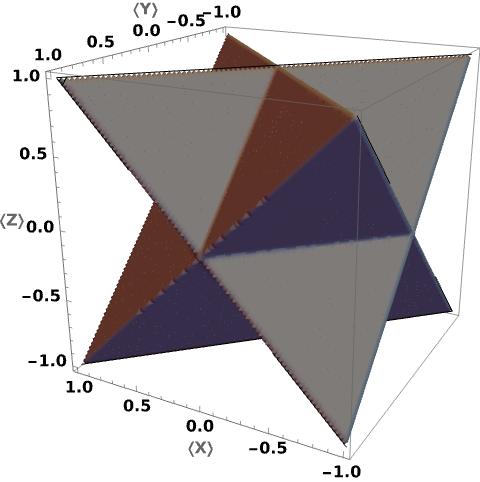}}\\
\subfloat[]{\includegraphics[width=0.4\textwidth]{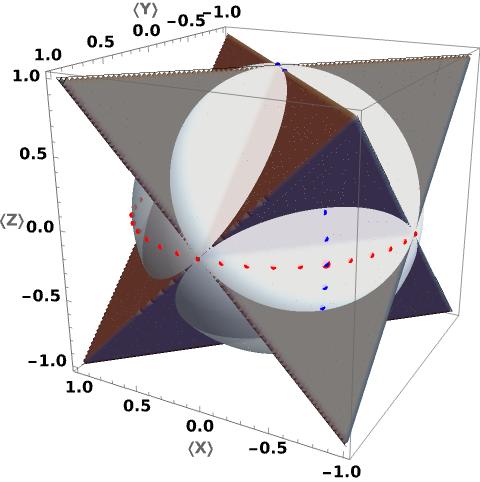}}
\caption{Bloch ball representation of qubit states $\rho$ (region inside the sphere), for which the product state $\rho\otimes\rho$ satisfies the inequalities (\ref{nonlinear}), with the given measurements \eqref{m1} and \eqref{m2}. (a) -- the set of values $\{\langle X \rangle,\langle Y\rangle, \langle Z\rangle\}$ which does not violate the inequality (\ref{nonlinear}). (b) -- the intersection of the region shown in (a) with a Bloch sphere. The points outside the contoured regions are pure quantum states violating all the inequalities (\ref{nonlinear}). We also plot the two families of experimentally tested states. The states $\ket{\psi} = \cos(\theta)\ket{0} + \sin(\theta)e^{i\phi}\ket{1}$ with fixed $\theta = \pi/4$ and $0\leq \phi \leq \pi$ are represented as red dots and the states with fixed $\phi = 3\pi/4$ and $0\leq \theta \leq \pi/2$ are represented as blue dots. The intersection between the two families is the state $\ket{\psi} = \frac{1}{\sqrt{2}}(\ket{0} + e^{i 3\pi/4} \ket{1})$.}
\label{f2}
\end{figure}

For the sake of simplicity, we consider qubit states prepared in a symmetric product state $\rho\otimes\rho$. The subset of such states satisfying inequality (\ref{nonlinear}) is shown in Fig. \ref{f2}.  



\subsection{Experiment}
An experimental test of the inequalities was performed on two $^{171}\text{Yb}^+$ ions in a linear Paul trap separated by about 5$\mu$m. The $^2S_{1/2}$ hyperfine `clock' states, $\ket{0} \equiv \ket{F=0,m_{F}=0}$ and $\ket{1} \equiv \ket{F=1,m_{F}=0}$, form an effective two-level system. 



Starting with the initial state $\ket{0}$ we apply a  single qubit rotation to each ion to prepare the state $|\psi\rangle \otimes |\psi\rangle$, where $\ket{\psi} = \frac{1}{\sqrt{2}}(\ket{0} + e^{i 3\pi/4} \ket{1})$ (see Methods). To measure Pauli observables (\ref{m1}) we then apply another single qubit rotation to transform the measurement basis from $Z$ to $X$ or $Y$ before detecting the ion state. To measure an expectation value of  $\langle \measA \measB \rangle$ a M{\o}lmer-Sorensen gate is applied to the pair of ions before the state detection. 

We observed strong violations of the four inequalities (\ref{eqn:exp_inequality_1st}-\ref{eqn:exp_inequality}), as shown in Table \ref{tab:table1}. Furthermore, the left-hand side of the inequality (\ref{nonlinear}) (evaluated for over 10000 measurements) is equal to $0.089(7) >0$, which exceeds the $0$ bound by more than 10 standard errors. Consequently, the system exhibits bi-contextuality.


\begin{table}[tb]
\caption{\label{tab:table1}%
Measurement outcomes for $\ket{\psi} = \ket{0} + e^{i 3\pi/4} \ket{1}$ are presented below. Results reported are experimental values with standard errors taken over more than 10000 measurements, not accounting for state preparation and measurements (SPAM) and gate fidelity errors. Strong violations are observed for all measurement settings since the measured values and upper bounds are all separated from each other by at least 10 measured value standard errors.
}
\begin{ruledtabular}
\begin{tabular}{lcdr}
\textrm{Inequality}&
\textrm{Measured}&
\multicolumn{1}{c}{\textrm{Upper}}&
\textrm{Lower}\\
\textrm{Type}&
\textrm{Value}&
\multicolumn{1}{c}{\textrm{Bound}}&
\textrm{Bound}\\
\colrule
$\langle \measA \measB \rangle/L1$ & -0.025(7) & -0.36(1) & -0.91(1)\\
$\langle \measA \measB \rangle/R1$ & -0.06(2) & -0.36(1) & -0.91(1)\\
$\langle \measA \measB \rangle/L2$ & -0.025(7) & -0.37(1) & -0.92(1)\\
$\langle \measA \measB \rangle/R2$ & -0.06(2) & -0.37(1) & -0.92(1)\\
\end{tabular}
\end{ruledtabular}
\end{table}

\begin{figure*}[t]
\includegraphics[width=0.99\textwidth]{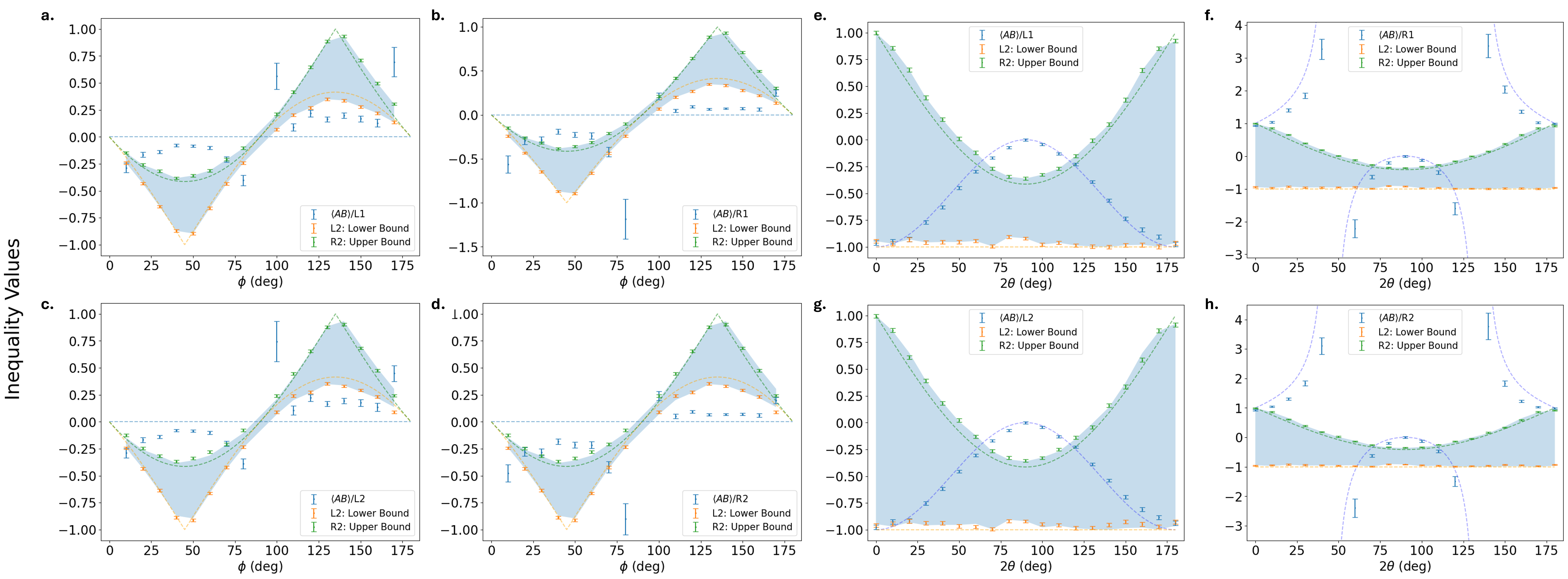}
\caption{Measurement outcomes for $\ket{\psi} = \cos{(\theta)} \ket{0} + \sin{(\theta)}  e^{i \phi} \ket{1}$. The dotted lines indicate theoretical expectations (see Methods). Markers indicate experimental values with standard errors over more than 5000 measurements, not accounting for state preparation and measurements (SPAM) and gate fidelity errors. The shadowed area indicates an experimental region whereby the inequality is satisfied. (a-d) The value of $\phi$ is varied from $0$ to $\pi$ with the value of $\theta$ set to $\pi/4$. However, the points corresponding to $0$, $\pi/2$, and $\pi$ have been omitted as they are too close to infinity points. We see the strongest violations near $\phi = \pi/4$ and $\phi = 3\pi/4$. (e-h) The value of $2\theta$ is varied from $0$ to $\pi$ with the value of $\phi$ set to $3\pi/4$. However, the points corresponding to $5\pi/18$ and $13\pi/18$ have been omitted from (f) and (h) as they are too close to infinity points. We see the strongest violations near $\theta = \pi/4$. } 
\label{fig:combined}
\end{figure*}

\begin{figure*}[t]
\includegraphics[width=0.99\textwidth]{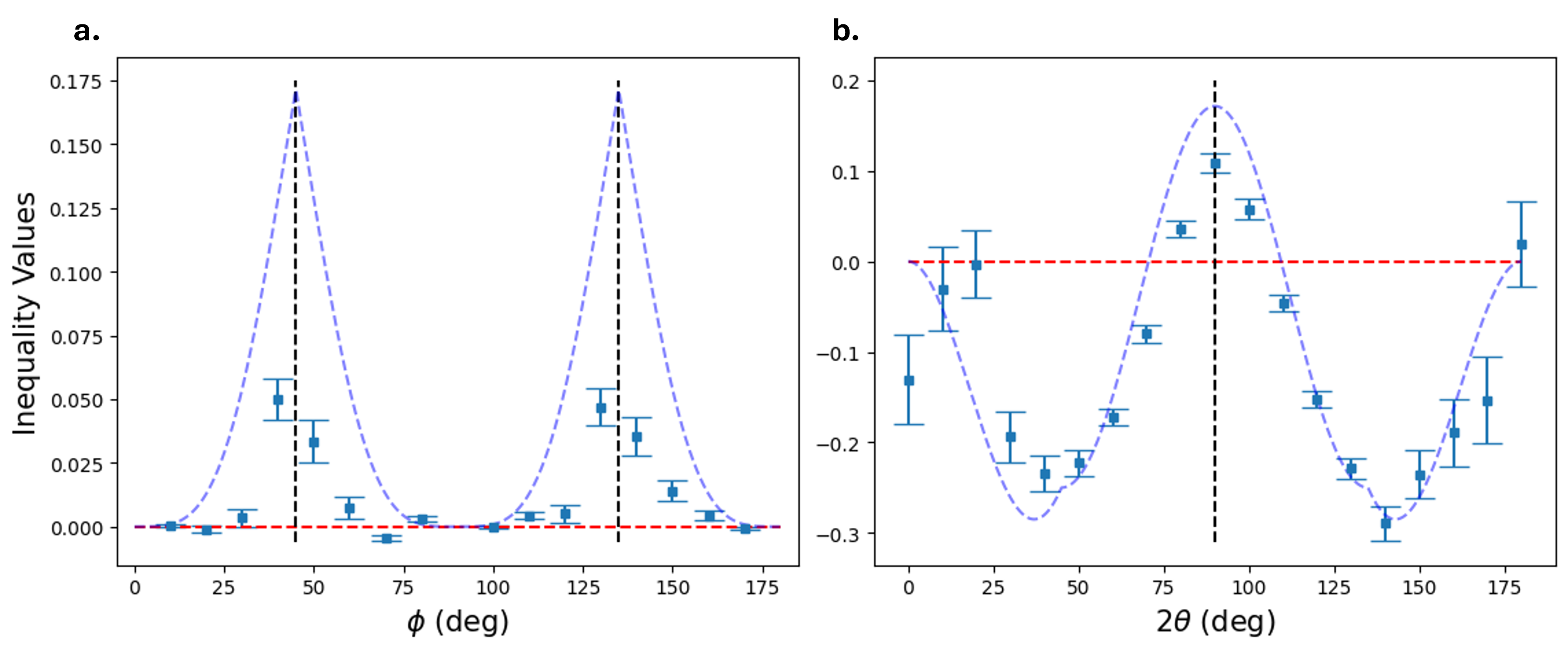}
\caption{Measurement outcomes for $\ket{\psi} = \cos{(\theta)} \ket{0} + \sin{(\theta)}  e^{i \phi} \ket{1}$ and the inequality (\ref{nonlinear}). The blue dotted lines indicate theoretical expectations (see Methods). The red dotted line represents the $0$ bound, above which the inequality is violated. Markers indicate experimental values with standard errors over more than 5000 measurements, not accounting for SPAM and gate fidelity errors. (a) The value of $\phi$ is varied from $0$ to $\pi$ with the value of $\theta$ set to $\pi/4$. However, the points corresponding to $0$, $\pi/2$, and $\pi$ have been omitted for reasons stated in Figure \ref{fig:combined}. We see the strongest violations near $\phi = \pi/4$ and $\phi = 3\pi/4$ represented by the black dotted lines. (b) The value of $2\theta$ is varied from $0$ to $\pi$ with the value of $\phi$ set to $3\pi/4$. We see the strongest violations near $\theta = \pi/4$ represented by the black dotted line.} 
\label{fig:combined eqn 41}
\end{figure*}

To further verify theoretical predictions, we performed additional measurements for state $|\psi\rangle \otimes |\psi\rangle$, where $\ket{\psi} = \cos \theta \ket{0} + e^{i \phi}\sin \theta \ket{1})$ and the angles $\theta$ and $\phi$ were varied. These results are plotted in Fig. \ref{fig:combined}. We observed good agreement with theoretical predictions. Furthermore, the same data set was used to plot the experimental values of the LHS of the inequality (\ref{nonlinear}) -- see Fig. \ref{fig:combined eqn 41}. 


\section{Discussion}

Our results have several implications, which we discuss below. 


\subsection{Preparation independence postulate}

The Preparation Independence Postulate (PIP) is a key assumption in multiple areas of quantum foundations, appearing in both the study of Ontological Models \cite{pusey2012} and Network Nonlocality \cite{tavakoli2022}. In ontological models, PIP plays a central role in the derivation of the Pusey-Barrett-Rudolph (PBR) theorem, whose conclusions no longer hold if this assumption is relaxed \cite{leifer2014}. In network nonlocality, PIP is a defining characteristic of bilocal models \cite{branciard2010}, although it has been shown that it is not always necessary to demonstrate quantum network nonlocality \cite{vsupic2020}.

While the assumption of independent preparations may seem reasonable when considering independently prepared quantum systems, its implications for ontological models—particularly noncontextual models—are more profound. It has been argued that PIP is equivalent to assuming local causality \cite{emerson2013}, and concerns have been raised regarding the composition of hidden variables, especially in scenarios involving entangling measurements \cite{schlosshauer2014}. Due to these criticisms, several works on ontological theorems have deliberately avoided assuming PIP \cite{ringbauer2015, barrett2014}.

A natural point of comparison arises between the role of PIP in the PBR theorem and in network contextuality scenarios. While both contexts rely on independent preparations, they target different aspects of hidden-variable models.

The PBR theorem considers two different preparations of a single system, each in a respective pure state $|\psi_a\rangle$ and $|\psi_b\rangle$. In an ontological model, these states correspond to probability distributions $\mu_a(\lambda)$ and $\mu_b(\lambda)$ over the ontic state space $\Lambda$. The theorem states that if $|\psi_a\rangle$ and $|\psi_b\rangle$ are non-orthogonal, a hidden-variable model where $\mu_a(\lambda)$ and $\mu_b(\lambda)$ overlap cannot fully reproduce quantum predictions. The argument relies on product-state preparations of multiple copies of the system, such as $|\psi_a\rangle\otimes |\psi_a\rangle$, $|\psi_a\rangle\otimes |\psi_b\rangle$, etc., and a single entangling measurement. Crucially, it assumes that these preparations correspond to factorizable probability distributions, $\mu_a(\lambda_1) \mu_b(\lambda_2)$, where $\lambda_i$ represents the ontic state of the $i$-th system.

By contrast, in network contextuality, we examine joint measurements on many independent systems rather than a single measurement on multiple copies of the same system. The prepared states, such as $\rho_a \otimes \rho_b$, need not be pure and can represent distinct physical systems, such as an electron's spin and a photon's polarization. The goal is to show that quantum measurement outcomes cannot be reproduced by noncontextual hidden-variable theories, which assume independent preparations correspond to factorizable distributions, $\mu_a(\lambda_1)\mu_b(\lambda_2)$.

In simpler terms, the PBR theorem leverages PIP to infer properties of the ontic states associated with a single system, while network contextuality uses PIP to analyze the structure of hidden-variable models for composite systems. Despite their differences, both cases highlight the tension between quantum predictions and the assumption that independent preparations correspond to statistically independent ontic states.

The results presented here reinforce the idea that PIP is not merely a strong assumption but, in some cases, an invalid one when attempting to reproduce quantum theory with noncontextual models. Our findings provide experimental support for the concerns raised in \cite{schlosshauer2014}, showing that constructing a noncontextual model for composite quantum systems requires more than simply combining noncontextual models for individual subsystems. As such, the independence of preparations in ontological models should be treated with caution, especially when making claims about the fundamental nature of hidden-variable theories.


\subsection{Certification of non-classicality of network nodes}

Our scheme offers a simple, device-independent method for certifying the non-classicality of a measurement apparatus in a laboratory setting. Specifically, if the bi-non contextuality inequality (Eq. (41)) is violated, then the measurement devices — labeled $\measA$ and $\measB$ in the third measurement context — must be non-classical. This conclusion is robust: the first two measurement settings serve to independently verify the statistical separation of the inputs to boxes $\measA$ and $\measB$, establishing the necessary independence assumptions.

This makes our scheme especially practical for verifying the integrity of quantum network nodes, such as those implementing entangling measurement gates. A user of such a node may request independently prepared qubit states from spatially separated sources and use them to test the measurement devices in situ. A violation of the inequality then serves as a certificate that the node performs genuinely non-classical operations — even in the absence of detailed knowledge about its internal structure.


\subsection{Bi-contextuality in super-quantum theories}

There is a set of bi-contextual correlations that cannot be realized within quantum theory. Consider the following set of averages obtained from measurements on two independently prepared systems:
\begin{eqnarray}
    & &\langle \alpha_1 \rangle = \langle \beta_2 \rangle = 1,~~~~\langle \alpha_2 \rangle = \langle \beta_1 \rangle = 0, \label{sq1} \\
    & &\langle \measA \rangle =  \langle \alpha_1 \rangle\langle \alpha_2 \rangle = 0,~~~~\langle \measB \rangle =  \langle \beta_1 \rangle\langle \beta_2 \rangle = 0, \label{sq2}
\end{eqnarray}
and
\begin{equation}\label{sq3}
    \langle \measA \measB \rangle =1.
\end{equation}
Note that in quantum case joint measurability of $\measA$ and $\measB$ is provided by the properties of complementary Pauli observables corresponding to $\alpha_i$ and $\beta_i$. However, for these Pauli observables Eqs. (\ref{sq1}) and (\ref{sq2}) imply $\langle \measA \measB \rangle =0$. On the other hand, non-negativity condition (\ref{nnc}) (see Methods) implies
\begin{equation}\label{cond}
 0=|\langle\alpha_i \rangle + \langle\beta_i \rangle|-1 \leq  \langle \alpha_i \beta_i\rangle \leq 1- |\langle\alpha_i \rangle - \langle\beta_i \rangle|=0,
\end{equation}
and
\begin{equation}
-1= |\langle \measA \rangle + \langle \measB \rangle|-1 \leq  \langle \measA \measB\rangle \leq 1- |\langle \measA \rangle - \langle \measB \rangle|=1.
\end{equation}
The averages (\ref{sq1}), (\ref{sq2}) and (\ref{sq3}) obey the above. However, they do not obey the preparation independence postulate
\begin{equation}
    \langle \measA \measB \rangle=\langle \alpha_1 \beta_1\rangle \langle \alpha_2 \beta_2\rangle, 
\end{equation}
because $\langle \measA \measB \rangle =1$ and (\ref{cond}) implies that $\langle \alpha_i \beta_i\rangle=0$.


\subsection{Non bi-contextual set is not convex}

Unlike standard set of non contextual behaviours, the set of  non bi-contextual behaviors is not convex. This is analogical to the set of network local behaviors, where PIP plays a key role. One can show this non-convexity via a simple example. Consider the following deterministic non bi-contexual behavior:
\begin{eqnarray}
    & &\langle \alpha_1 \rangle = \langle \beta_2 \rangle = 1,~~~~\langle \alpha_2 \rangle = \langle \beta_1 \rangle = 1, \label{d11} \\
    & &\langle \measA \rangle =  \langle \alpha_1 \rangle\langle \alpha_2 \rangle = 1,~~~~\langle \measB \rangle =  \langle \beta_1 \rangle\langle \beta_2 \rangle = 1, \label{d12}
\end{eqnarray}
and
\begin{equation}\label{d13}
    \langle \measA \measB \rangle =1.
\end{equation}
Next, consider another deterministic non bi-contexual behavior:
\begin{eqnarray}
    & &\langle \tilde\alpha_1 \rangle = \langle \tilde\beta_2 \rangle = 1,~~~~\langle \tilde\alpha_2 \rangle = \langle \tilde\beta_1 \rangle = -1, \label{d21} \\
    & &\langle \tilde \measA \rangle =  \langle \tilde\alpha_1 \rangle\langle \tilde\alpha_2 \rangle = -1,~~~~\langle \tilde \measB \rangle =  \langle \tilde\beta_1 \rangle\langle \tilde\beta_2 \rangle = -1, \label{d22}
\end{eqnarray}
and
\begin{equation}\label{d23}
    \langle \tilde \measA \tilde \measB \rangle =1.
\end{equation}
An even mixture of these two behaviors leads to the super-quantum bi-contextual behavior (\ref{sq1}-\ref{sq3}).  


\subsection{Relation to Bell scenario}

Our scenario can be seen as a \textit{reversed} Bell scenario. In a standard Bell scenario, a single source emits a pair of systems, each of which travels to a different laboratory where it is measured. These laboratories are space-like separated, and the choice of measurements is made independently and randomly within each laboratory. As a result, the measurements performed on the two systems are independent. The outcomes of these measurements may not admit a classical description if the correlations between the systems -- established at the source -- are nonclassical. In particular, if the systems are quantum and entangled, there exist numerous strategies for measuring them to observe quantum nonlocality \cite{brunner2014}. Quantum nonlocality is a fundamental phenomenon that relies on entanglement as a resource. In contrast, local measurements on systems prepared in a product (non-entangled) state do not lead to quantum nonlocality. From this perspective, entangled states are resourceful, while product states are resourceless.

In our scenario, we encounter an inverted situation. This time, two independent sources each send a system into a single laboratory, where the systems are jointly measured. We show that these measurements may not admit a classical description. In particular, if the systems are quantum, their independent preparation implies that they are in a product state. Therefore, from the perspective of our scenario, product states are the resourceful ones, as they give rise to the phenomenon of bi-contextuality. Conversely, entangled states do not appear in our scenario at all, as they cannot be created by independent sources. Based on this observation, we argue that bi-contextuality complements nonlocality, as it enables the detection of nonclassicality in a completely different and disjoint set of states.


\subsection{Relation to Mermin-Peres square}

One of the most well-known proofs of state-independent quantum contextuality was provided by Mermin and Peres \cite{peres1990, mermin1990} and is known as the Mermin-Peres square. It consists of nine two-qubit operators arranged in a $3\times 3$ square table:
\begin{center}
\begin{tabular}{ |c|c|c| } 
 \hline
 $X\otimes \openone$ & $\openone \otimes X $ & $X\otimes X$ \\ 
  \hline
$\openone\otimes Y$ & $Y \otimes \openone $ & $Y\otimes Y$ \\ 
  \hline
$X\otimes Y$ & $Y\otimes X$ & $Z\otimes Z$ \\ 
 \hline
\end{tabular}
\end{center}
where $X$, $Y$ and $Z$ are Pauli operators and $\openone$ is the identity operator. Each row and column contains three mutually commuting operators, meaning they can be measured jointly. Furthermore, each of the nine operators has two possible eigenvalues, $+1$ and $-1$. The contextuality of this set arises from the fact that no pre-assigned outcome ($\pm 1$) can satisfy the multiplication constraints imposed by the operators. Specifically, the product of the three operators in rows 1, 2, and 3, as well as in columns 1 and 2, yields the identity operator $\openone$, while the product of the three operators in column 3 yields $-\openone$. If one were to multiply the pre-assigned outcomes of all nine observables row by row, the result would be $+1$, whereas multiplying them column by column would yield $-1$ a contradiction. Crucially, this impossibility arises solely from the algebraic structure of the operators, independent of any particular two-qubit state.

Although the full set of nine operators provides a proof of state-independent contextuality, certain subsets exhibit state-dependent contextuality (or non-locality). In particular, considering only the measurements from rows 1 and 2 and columns 1 and 2 yields a Clauser-Horne-Shimony-Holt (CHSH) Bell test \cite{CHSH}, which reveals non-locality for specific entangled two-qubit states. Here, we identify yet another subset of the Mermin-Peres square that serves as a bi-contextuality test: columns 1 and 2, along with row 3. This finding establishes a further connection between contextuality and bi-contextuality, reinforcing our observation that bi-contextuality complements non-locality.


\section{Methods}

\subsection{Non-negativity condition}

Consider two binary random variables, $Q$ and $R$, with the corresponding outcomes $q,r=\pm 1$. Their joint probability distribution can be represented in the following form
\begin{equation}\label{joint}
p(q,r) = \frac{1}{4}\left(1+q\langle Q\rangle + r\langle R\rangle + qr \langle QR\rangle\right),     
\end{equation}
where we use a shorthand notation $p(Q=q,R=r)=p(q,r)$. Non-negativity of the above distribution implies
\begin{equation}\label{nnc}
 |\langle Q \rangle + \langle R \rangle|-1 \leq  \langle QR \rangle \leq 1- |\langle Q \rangle - \langle R \rangle|.
\end{equation}

\subsection{Derivation of bi-contextuality inequality}

We recall that if there exists a non-bi-contextuality model, then $\alpha_i$ and $\beta_i$ are jointly distributed. Such distribution must take form (\ref{joint}). The non-negativity of probability implies the following measurable bounds on the unmeasurable correlation $\langle \alpha_i \beta_i\rangle$ ($i=1,2$)
\begin{equation}
 |\langle\alpha_i \rangle + \langle\beta_i \rangle|-1 \leq  \langle \alpha_i \beta_i\rangle \leq 1- |\langle\alpha_i \rangle - \langle\beta_i \rangle|.
\end{equation}
 To simplify notation let us define
 \begin{eqnarray}
     L_i &=& |\langle\alpha_i \rangle + \langle\beta_i \rangle|-1  \\
     R_i &=& 1- |\langle\alpha_i \rangle - \langle\beta_i \rangle|.
 \end{eqnarray}

\begin{figure}[t]
\includegraphics[width=0.5\textwidth]{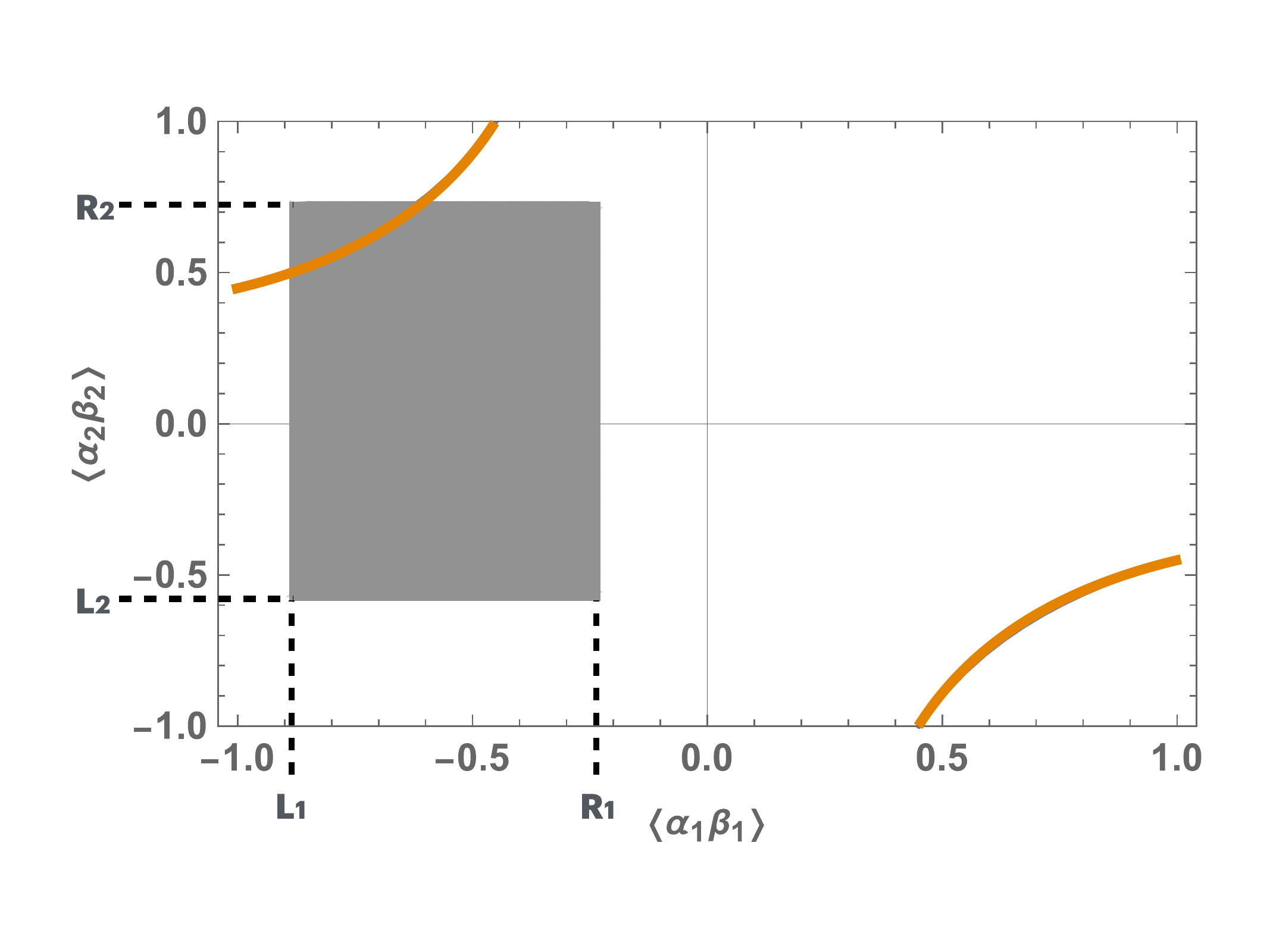}
\caption{Graphical representation of the unmeasurable correlations $\langle \alpha_1\beta_1\rangle$ and $\langle \alpha_1\beta_1\rangle$. The allowed region is represented by the gray rectangle defined by bounds $L_i$ and $R_i$ ($i=1,2$). The relation $\langle \measA \measB \rangle = \langle \alpha_1 \beta_1\rangle\langle \alpha_2 \beta_2\rangle$ is represented by the orange hyperbola. The non-bi-contextual model exists whenever the orange hyperbola passes through gray rectangle.} 
\label{f3}
\end{figure}

Next, recall that non-bi-contextuality model also implies
\begin{equation}
    \langle \measA \measB \rangle = \langle \alpha_1 \beta_1\rangle\langle \alpha_2 \beta_2\rangle,
\end{equation}
which leads to a hyperbolic relation between $\langle \alpha_1 \beta_1\rangle$ and $\langle \alpha_2 \beta_2\rangle$, see Fig. \ref{f3}.

The model exists if the hyperbola passes through the region defined by the bounds \(L_i\) and \(R_i\). These bounds define a rectangular region in the \(\langle \alpha_1\beta_1 \rangle\)–\(\langle \alpha_2\beta_2 \rangle\) plane, and the hyperbola intersects this region if it crosses any of its sides:

\begin{equation}
    L_2 \leq \frac{\langle \measA \measB \rangle}{L_1} \leq R_2, \tag{left side}
\end{equation}
\begin{equation}
    L_2 \leq \frac{\langle \measA \measB \rangle}{R_1} \leq R_2, \tag{right side}
\end{equation}
\begin{equation}
    L_1 \leq \frac{\langle \measA \measB \rangle}{R_2} \leq R_1, \tag{upper side}
\end{equation}
\begin{equation}
    L_1 \leq \frac{\langle \measA \measB \rangle}{L_2} \leq R_1. \tag{lower side}
\end{equation}

The model does not exists if all of the above inequalities are violated. Finally, note that these inequalities are expressed via measurable quantities, hence they can be tested in an experiment. 

There is a simpler, yet only necessary, condition for a non-bi-contextual model. Note that the unmeasurable correlations obey
\begin{equation}
   |\langle \alpha_i \beta_i\rangle|\leq 1. 
\end{equation}
In addition
\begin{equation}
  \frac{|\langle \measA \measB\rangle|}{\max\{|L_2|,|R_2|\}} \leq |\langle \alpha_1 \beta_1\rangle|. 
\end{equation}
and
\begin{equation}
  \frac{|\langle \measA \measB\rangle|}{\max\{|L_1|,|R_1|\}} \leq |\langle \alpha_2 \beta_2\rangle|. 
\end{equation}
Therefore,
\begin{equation}
    |\langle \measA \measB\rangle| \leq \min_{i=1,2}\{\max\{|L_i|,|R_i|\}\}.
\end{equation}
If the above inequality is violated, then the measurements on the system are bi-contextual, but if it is not, then it is not guaranteed that the non-bi-contextual model exists.

\subsection{Single inequality}

Here we present a derivation of the inequality (\ref{nonlinear}), which succinctly combines all the other inequalities derived in the paper.

We know that $\langle\alpha_i\beta_i\rangle=\lambda_iL_i+(1-\lambda_i)R_i$, where $0\leq\lambda_i\leq 1$. This stems from the positivity condition of the probability distribution of the variables $\alpha_i,\beta_i$. Their independence requires that $\langle \measA \measB\rangle=\langle\alpha_1\beta_1\rangle\langle\alpha_2\beta_2\rangle$. This is a function of two independent variables $\lambda_1$ and $\lambda_2$ whose domain is a square. The maximum, $max$, and the minimum, $min$, are located in the four corners of the square and, obviously, $min\leq \langle \measA \measB\rangle \leq max$. This is equivalent to the inequality (\ref{nonlinear}).

\subsection{Experimental Details}
The experimental platform consists of a linear radio-frequency (RF) Paul trap with 4 rods and 2 needle electrodes \cite{PhysRevA.76.052314}. An RF signal at 17.7 MHz is applied to one diagonal pair of rods, while DC voltages are sent to the other pair of rods and both needles. Within the Paul trap, we confine two $^{171}\text{Yb}^+$  ions and encode quantum information within the hyperfine `clock' states $|0\rangle \equiv |F=0,m_{F}=0\rangle$ and $|1\rangle \equiv |F=1,m_{F}=0\rangle$. These hyperfine qubit states exist in the $^{2}\text{S}_{1/2}$ electronic ground state and are separated by 12.6 GHz \cite{PhysRevA.76.052314}. The transition between $^2S_{1/2}$ and $^2P_{1/2}$ at 369.52 nm is used to perform Doppler \cite{RevModPhys.75.281} and electromagnetically-induced transparency (EIT) cooling \cite{PhysRevLett.125.053001}, optical pumping, and state detection. With two ions, there are six different normal modes of motion, split into three center-of-mass modes and three out-of-phase modes, each split into three spatial dimensions. The frequencies of these six modes range from approximately 0.6 MHz to 1.5 MHz with our axial center-of-mass mode frequency at $\nu_{AxCOM}=0.62 \text{MHz}$. 

Our qubit states are coupled with coherent stimulated Raman transitions implemented with a 355nm mode-locked pulsed laser split into two beams. These two Raman beams propagate into the vacuum chamber with an angle of $135^{\circ}$ relative to each other. The beat note between the two Raman beams drives coherent rotations within our qubit Bloch sphere. The two global Raman arms are aligned to the trap such that both qubits have equal Rabi frequencies. 

An experimental run begins with the two ions optically pumped to the $\ket{0}$ state, and cooled to the motional ground state ($\langle\bar{n}\rangle \approx 0.1$) using Doppler cooling \cite{RevModPhys.75.281}, EIT cooling \cite{PhysRevLett.125.053001}, and resolved sideband cooling \cite{RevModPhys.75.281}. The initial state $\ket{\psi} = \ket{0} + e^{i 3\pi/4} \ket{1}$ is then prepared using a $\pi/2$ rotation pulse ($\tau_{\pi/2} \approx 5 \; \mu\text{s}$) (which can be tuned via the duration of the stimulated Raman transition) and by imparting the $e^{i 3\pi/4}$ phase factor using an appropriate phase difference between the Raman beams. The ions in the trap are separated by about 5$\mu$m and do not interact with each other during the state preparation. 

The sole measurable quantity in this set-up is the population of the two qubit states, corresponding to projections to the $Z$ eigen-states. To measure the expectation values of other Pauli operators, the initial state is rotated to the corresponding basis. These are evaluated as: 
\begin{eqnarray}
    \bra{\psi}X\ket{\psi} = \bra{\psi}R_Y^\dagger\bigg(\frac{\pi}{2}\bigg)\; Z \; R_Y\bigg(\frac{\pi}{2}\bigg) \ket{\psi} \\
    \bra{\psi}Y\ket{\psi} = \bra{\psi}R_X^\dagger\bigg(\frac{\pi}{2}\bigg)\; Z \; R_X\bigg(\frac{\pi}{2}\bigg) \ket{\psi} 
\end{eqnarray}
Here, $R_{X,Y}(\theta)$ are single qubit rotations in the labeled axis by angle $\theta$ and the Bloch sphere axes $X$ and $Y$ can be selected by selecting a $\pi/2$ relative phase difference in one of the Raman arms. For the joint measurements,
\begin{eqnarray}
    \bra{\psi \otimes \psi} (XY \otimes YX) \ket{\psi \otimes \psi} \nonumber \\
    = \bra{\psi \otimes \psi} \text{MS}^\dagger (Z \otimes Z) \text{MS} \ket{\psi \otimes \psi}
\end{eqnarray}
Here, MS refers to the M{\o}lmer-Sorensen gate, the native two-qubit interaction in a trapped ion system ($\tau_{\text{MS}} \approx 250 \mu$s). It is implemented by illuminating both ions with bichromatic laser beams that simultaneously drive both the red-sideband (RSB) and blue-sideband (BSB) Raman transitions. The RSB(BSB) transitions subtract(add) a quanta of motional energy from a single motional mode via a flip in the qubit state. The resultant bichromatic waveform's amplitude is modulated further with a $\sin^2$ envelope \cite{PhysRevLett.123.260503}. For this experiment, we implement the MS gate using a radial out-of-phase mode with a mode frequency of approximately 0.79 MHz. 

After these secondary rotations, a state detection beam illuminates the ions for $ 1$ms. The scattered photons are collected and relayed onto a multi-channel photomultiplier tube (MCPMT). This allows for simultaneous detection of both ions' excited state populations, where a collection of $<2$ photons indicates a qubit in $\ket{0}$, and $\ket{1}$ otherwise. 

The typical state detection error in our setup is $2.6(1)\%$, limited mostly by the ion fluorescence collection efficiency, and the typical state preparation error (to the $\ket{0}$ qubit state) is $2.3(1)\%$. The fidelity of the MS gate (without the SPAM error subtraction) is $91(2)\%$. The MS gate calibration and fidelity measurement procedures follow~\cite{akerman_universal_2015}.


Theoretical simulations of the experimental results were performed using the QuTiP quantum computing framework. The two-ion system can be represented in QuTiP as a tensor of two qubits. The arbitrary qubit rotation gate can be manually coded as a quantum object class and the MS interaction can be imported via the qutip-qip package. 

\subsection{Theoretical Bounds for Varying Phase \& Amplitude}
Consider a general spin-$\frac{1}2$ state, $\ket{\psi} = \cos(\theta)\ket{\downarrow} + \sin(\theta)e^{i\phi}\ket{\uparrow}$. The corresponding expectation values are:
\begin{eqnarray}
    \braket{\psi | X | \psi} = \sin(2\theta) \cos(\phi) \nonumber \\
    \braket{\psi | Y | \psi} = \sin(2\theta) \sin(\phi)
\end{eqnarray}
For the measurement settings
\begin{eqnarray}
    \alpha_1 = X \otimes \openone, \beta_1 = Y \otimes \openone, \alpha_2 = \openone \otimes X, \beta_2 = \openone \otimes Y \nonumber
\end{eqnarray}
The lower and upper bounds for bi-contextuality are 
\begin{eqnarray}
    L_1 = L_2 = |\sin(2\theta) [\cos(\phi) + \sin(\phi)]| - 1 \nonumber \\
    R_1 = R_2 = 1 - |\sin(2\theta) [\cos(\phi) - \sin(\phi)]|
\end{eqnarray}
And the joint measurement outcome
\begin{eqnarray}
    \measA \measB = \langle Z \otimes Z \rangle = \cos^2(2\theta)
\end{eqnarray}

\section*{Acknowledgements}

We would like to thank P. B{\l}asiak, R. Chaves, A. Grudka, M. Karczewski, J. Stempin, M. Terra Cunha, A. W{\'o}jcik, J. W{\'o}jcik, for stimulating discussions and helpful comments. 

PK was supported by the Polish National Science Centre (NCN) under the Maestro Grant no. DEC-2019/34/A/ST2/00081 and by the CAPES/PRINT of Photonics Applied to Communication and Information Theory. GR was supported by the São Paulo Research Foundation (FAPESP), Brasil (Process Numbers 2021/10548-0 and 2023/04053-3). RR is supported by the Brazilian National Council for Scientific and Technological Development (CNPq) (INCT-IQ and Grant Number 316657/2023-9). This project is supported by the National Research Foundation, Singapore through the National Quantum Office, hosted in A*STAR, under its Centre for Quantum Technologies Funding Initiative (S24Q2d0009) and Quantum Engineering Programme (W21Qpd0208). 


\bibliographystyle{apsrev4-2}
\bibliography{apssamp.bib}


\end{document}